\documentclass[11pt]{article} 
\usepackage{moriond,epsfig} 
\def\be{\begin{equation}} 
\def\ee{\end{equation}} 
\def\bea{\begin{eqnarray}} 
\def\eea{\end{eqnarray}} 
 
\begin{document} 

{\it ~~~~~ Proceedings of the V. Rencontre de Moriond in Mesoscopic Physics 2004 }
\\ \\
\title{  Distribution of the Kondo Temperature in Mesoscopic
  Disordered Metals   } 
 
\author{\underline{Stefan Kettemann}$^1$ } 
 
\address{$^1$ I. Institut f\"ur 
  Theoretische Physik, Universit\" at Hamburg, Jungiusstra\ss{}e 9, 20355 
  Hamburg, Germany }

\maketitle
\abstracts{
  The   Kondo temperature  of  a magnetic impurity 
  in a  weakly disordered metal is  distributed due 
    to  the randomness in the   
 local exchange coupling, and the  local electronic  density of
 states (LDOS).  
  We show that in a
 closed, phase coherent  metal particle  
 the resulting distribution of $T_K$ is strongly asymmetric and scales
 with the mean level spacing $\Delta$. 
 Its    width  is 
  $2 \sqrt{\Delta/\beta T_K}$, where $\Delta$ is the  
 mean level spacing, and  $\beta=1,2$, 
 with, without time reversal symmetry, respectively. 
 Increasing the density of magnetic impurities, the distribution of Kondo  
 temperatures (DKT) is   found to become more narrow.  
 Corrections to these results  
 due to  Anderson localisation are discussed.  
} 
 
\section{Introduction} 

 A  local magnetic impurity is known to  change 
 the ground state of a Fermi liquid due to  correlations 
 created by the exchange interaction between its localised spin 
 and the delocalised electrons. 
 As a result, the magnetic impurity spin is screened  at zero temperature 
 by the     formation of the Kondo  singlet with  the conduction
 electrons.  
  Disorder 
 is affecting the formation of the Kondo singlet 
 in various ways. 
 The distribution of exchange couplings, due to 
 random positioning of the magnetic impurity in the host lattice,  
  directly results in a corresponding distribution of the Kondo
 temperature \cite{bhatt,langenfeld}. However,   
 the distribution of  LDOS   
 in disordered metals is related nontrivially through an integral 
 equation with the Kondo temperature.
 In a previous work 
 the  distribution of the Kondo temperature (DKT)
  has been  related directly to the distribution of
 the LDOS 
 at the Fermi energy \cite{kotliar,miranda}. 
 This  could have some justification in open systems \cite{mirlin}. 

 Here we rather consider a magnetic impurity 
 in a closed, phase coherent metal particle, 
 where the  energy levels $E_l$ are discrete, a Kondo box \cite{kroha,suga}.  
  Then, the  self consistent equation, determining the Kondo temperature  
 of a magnetic impurity at position ${\bf r}$  
 is \cite{nagaoka} 
\begin{equation} \label{tk} 
1 = J({\bf r}) Vol. \int d \epsilon \frac{\rho (\epsilon, {\bf r})}{\epsilon}  
\tanh \left( \frac{\epsilon}{2 T_{\rm K}} \right).  
\end{equation} 
 The local density of states 
$ \rho (\epsilon, {\bf r})$ 
 is   in terms  
 of the local wave function amplitudes $\psi_l ( {\bf r})$ 
 of eigenstates $\mid l >$ of the disordered Hamiltonian  
 $H$  given  by 
\begin{equation} 
\rho  (\epsilon, {\bf r}) =  \sum_l \mid\psi_l ( {\bf r})\mid^2 
\delta \left( \epsilon - E_l \right), 
\end{equation} 
 Thus, Eq. (\ref{tk}) can be rewritten as  
\begin{equation} 
1= \frac{1}{x} \sum_{l=1}^N \frac{x_l}{s_l} \tanh \left(  
\frac{s_l}{2 \kappa} 
\right), 
\end{equation} 
  where  
$ x= \Delta/J$, where $\Delta = 1/\nu Vol.$. 
 $\nu$ is the average density of states 
 which is $\nu = m/2 \pi \hbar^2$ per spin channel  
 in a 2D electron system (2DES). $N = E_F/\Delta$, with the Fermi energy
  $E_F$  and $\kappa = T_{\rm K}/\Delta$.  
$x_l = Vol.  \mid\psi_l ( {\bf r})\mid^2$ is the  
local  eigenfunction probability.  
$s_l = E_l/\Delta$ are the Eigen energies in units of $\Delta$
 as measured relative to  Fermi energy $E_F$.

\section{ Distribution of $T_K$ }  
 
  In a disordered, phase coherent  
 metal particle, the electron levels  
 repel each other, and the level spacings have  the  
 Wigner-Dyson distribution \cite{mehta}.  
  In this regime, the  square of the wave function amplitudes  
 $x_l$ and the Eigenenergies $s_l$, are distributed independently.  
$x_l$ obeys the  exponential  Porter--Thomas distribution,  
 which is in the unitary regime, with broken  
 time reversal invariance  (GUE) and in the time reversal symmetric
 regime (GOE), 
 given respectively by \cite{mirlin} 
\begin{equation} 
P_{\rm GUE}(x_l) = \exp ( - x_l), ~~~~~~~~~~~~~~~~~~~ P_{\rm GOE} (x_l) = \frac{1}{\sqrt{2 \pi x_l}} \exp \left( - \frac{x_l}{2} \right). 
\end{equation} 
 The probability to find the l-th 
 energy level above the Fermi energy at energy  $s_l$ can be obtained from
 an integration over the measure of  
 random matrix theory \cite{mehta}. 
 For large $l \gg 1$ 
these probabilities are in good approximation  Gaussian
 distributed  
 around the equally spaced level spectrum $l$, $l = 1,2,...$
 with widths which scale like $\sqrt{l}$ and depend on the symmetry
 of the Hamiltonian. 
 For broken  
 time reversal invariance (GUE), one obtains: 
\begin{equation}
P_{\beta} (s_l) \approx \exp \left( - c_{\beta} \frac{(s_l -l)^2}{l} \right),
\end{equation}
where $ c_\beta \approx  \pi/4, 4/\pi$, for $\beta=1$, GOE, $\beta=2$, GUE, 
 respectively. 
  The distribution of  exchange couplings $x$,  
 $P_J(x)$ is a   
 function of the random spatial position of the magnetic impurity in the  
 host lattice \cite{langenfeld}, and  
 can thus be taken to be  independent from the distribution of the  
 conduction electron probabilities $x_l$.  We will therefore take 
 $x$ to be fixed for the moment. 
 The distribution of Kondo temperatures $T_{\rm K}$  
 can thus   in the random matrix theory regime be expressed as,   
\begin{eqnarray} \label{general} 
 P(T_{\rm K}) = \int   \prod_l d x_l P(x_l) 
\prod_l d s_l P(s_l)  
\delta \left( 1 -  
\sum_{l=1}^{N} \frac{x_l}{x s_l} 
 \tanh \left[ \frac{s_l}{2 \kappa} \right] \right) 
\mid  \partial_{T_{\rm K}} \sum_{l=1}^{N} \frac{x_l}{x s_l}  
\tanh \left[ \frac{s_l}{2 \kappa} \right]\mid.   
\end{eqnarray}

Approximating $\tanh ( s_l/2\kappa ) = 1$ for $ s_l \ge \kappa$ and  
$\tanh ( s_l/2\kappa) = s_l/2\kappa$ for $ s_l \le \kappa$,  
 we can perform the integrals over $x_l$. 
We find that   the energy levels with $s_l < \kappa$ 
  enter the distribution  
 only through the number $n_K$ of the energy levels which  
 are in the interval  $0 \le s_l \le T_K$, $l=1,2,...,n_K$. 
 Thus,  for $\kappa > 1$, 
 the random distribution of the energy levels
 enters the expression for the DKT only 
 through  
 the distribution function of $n_K$. 
 For $ N \rightarrow \infty$ 
 one can evaluate the resulting 
 expression in the two limits $n_K \gg1$ and 
$n_K =1$. 
  One obtains 
 the DKT 
   in  good approximation in both limits and interpolates 
 the distribution  for finite $n_K$, 
\begin{eqnarray} 
 P (T_{\rm K})=\frac{K_{\beta} \Delta}{T_{\rm K}^2} \sum_{n_K=1}^{\infty}\exp\left[ - \frac{c_{\beta}}{n_K} \left(\frac{T_K}{\Delta} - n_K \right)^2-\frac{\beta n_{\rm K}}{8}\left(-\frac{n_K\Delta}{2 T_K} + \ln \frac{n_K }{\kappa_0}  
\right)^2 \right]\left(n_K-1+e^{-\frac{n_K\Delta}{T_K}}\right),
\end{eqnarray}  
 where $K_{\beta} = \sqrt{\pi c_{\beta}} /2 $, $\kappa_0 = D \exp ( - x)/\Delta$. 
 Thus,  the Kondo temperature $T_K$  has a very asymmetric distribution, and 
 scales with the energy level spacing  $\Delta$.   
While 
 the  average value of the Kondo temperature is  
 found to be  independent of the symmetry of the  
 Hamiltonian,  $T_K^0$,  
  the  width  
 is  larger with unbroken time reversal symmetry,  
\begin{equation} 
\Gamma_{\beta} =  2 \sqrt{\frac{2 \Delta}{\beta T_K}}.   
\end{equation}  
It  vanishes in the infinite volume limit, $\Delta \rightarrow 0$,  
 suggesting that  in a large metallic 
 sample the distribution of the Kondo temperature 
 is only due to  local fluctuations of the  
 exchange couplings $ J/\Delta = 1/x$ \cite{bhatt,langenfeld}.  
 When there is  time reversal symmetry  
 the energy level repulsion  is weaker  and the tendency to localisation  
  stronger. As a consequence  the probability of a 
 vanishing wave function at a position ${\bf r}$ is enhanced,  
 as well as the probability of large wave function splashes.  
 This explains that the 
  DKT is wider  than in the unitary regime.  
 
\begin{figure}
\begin{center}
  \includegraphics[width=.55 \textwidth ]{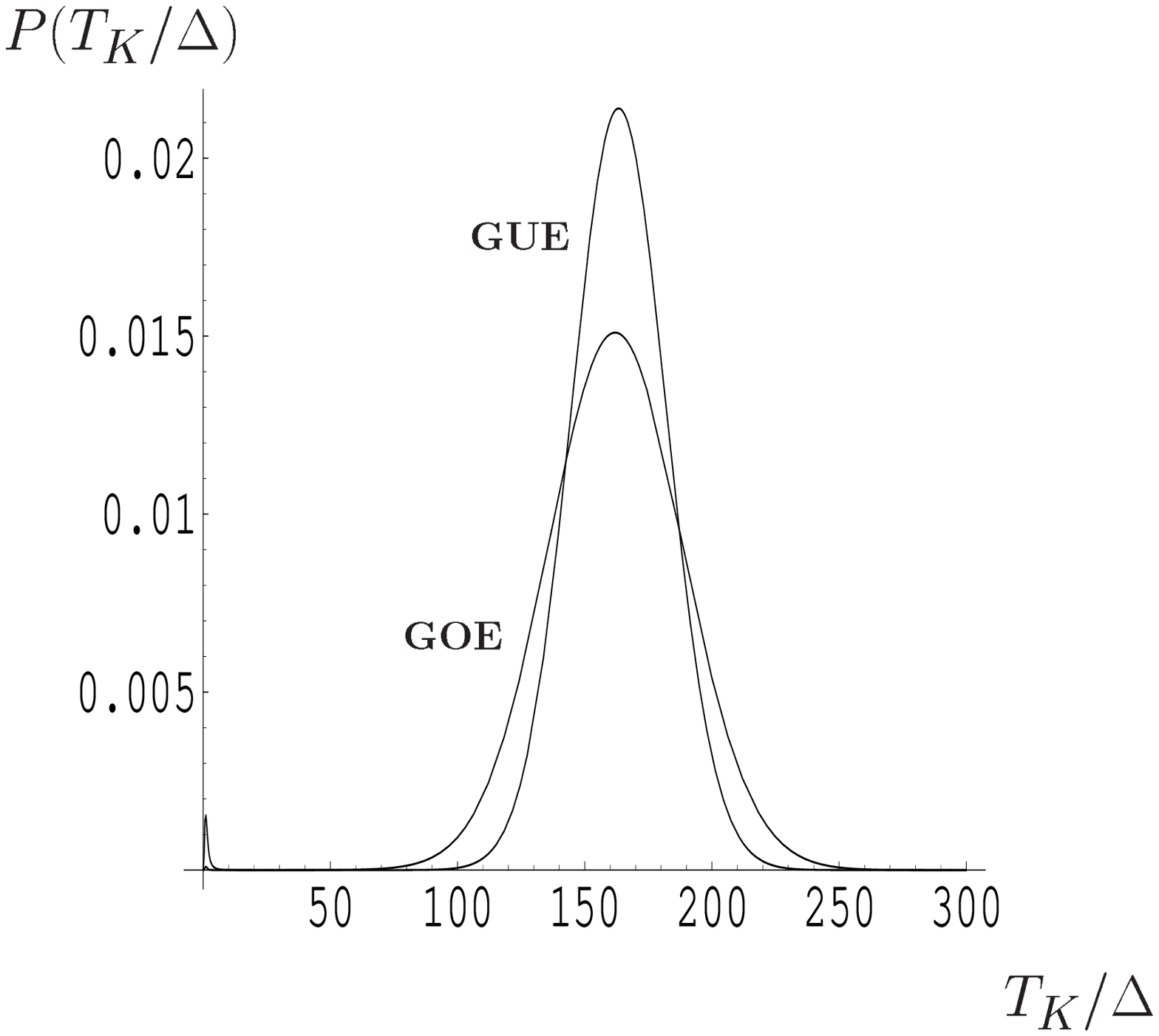}
\vspace{-5.5cm}
\caption{ The distribution of the Kondo temperature  $T_K$ as scaled
 with the level spacing $\Delta$ for the GOE and GUE ensemble. 
   }
 \label{loclength}
\end{center}
\vspace{-.5cm}
\end{figure}
\section{ Dependence of the Distribution of the Kondo Temperature  
on  the  Concentration of magnetic impurities }  
 
  Without an external magnetic field, $B =0$,  
 the crossover between the orthogonal and the unitary regime  
 is in a mesoscopic sample of size $L$, where the phase coherence  
 length $L_{\varphi}$ exceeds $L$, governed by the parameter \cite{hikami} 
\begin{equation} 
X_s = \frac{L^2}{D \tau_s},  
\end{equation} 
 with the diffusion constant $ D = v_F l/d$, where $d$ is the dimension 
of diffusion, and $l$ the elastic mean free path. 
 The spin scattering rate $1/\tau_s$ is renormalised by the  
 Kondo correlations with a maximal value limited  by  the unitary limit 
 of the scattering crossection, yielding in 2 dimensions, 
\begin{equation} 
\frac{1}{\tau_{s M}} =  n_M \frac{v_F}{ 2 k_F},  
\end{equation} 
where $n_M =1/R^2$ is the concentration of magnetic impurities 
 with $R$ the distance between them.  
 Thus, the maximal value of the crossover parameter is given in 2D by,  
\begin{equation} 
X_{s M} = \frac{N_M}{g},  
\end{equation} 
 where $N_M = n_M L^2$ is the number of magnetic impurities in the sample 
 of size $L$, and $ g = k_F l \gg 1$ is the dimensionless
 elastic mean free path in the metal.    
 When there are only few  magnetic impurities,  
 $N_M < g$, the crossover paramter $X_{s} <  1$ is small,   
 and the metallic particle is  in the orthogonal regime.  
 Increasing the concentration of magnetic impurities,  
 the parameter $X_s$ increases. 
 As found in the previous section this is accompanied by an  
 according decrease  of the width of the  DKT.   
 The spin scattering rate
  scales for   $\Delta \ll T_K$ like  $(\Delta/T_K)^2$ \cite{nozieres}. Thus, 
\begin{equation}
X_s = \frac{N_M \Delta^2}{g <T_K^2>},
\end{equation}
 and we conclude that the distribution $P(T_K)$ follows
 for $ N_M < g <T_K^2>/\Delta^2$ the wider orthogonal distribution, 
 crossing over to the unitary distribution in the opposite 
 regime, $ N_M > g <T_K^2>/\Delta^2$, where $<...>$ denotes the 
 average taken  over the $N_M$ magnetic impurities. 

 The superexchange between magnetic impurities  
 competes with the Kondo screening.  
 When $J_{RKKY}$ exceeds the Kondo temperature $T_K$  
 of single magnetic impurities, the spin of  magnetic impurities  
 is quenched, forming an array of  classical spins, 
 whose spin scattering rate 
 is by factor $1/3$ smaller than for free 
 spins \cite{spinglass}.   
 Thus the question arises, if the  unitary regime of the  
 distribution of Kondo temperatures  
 can be reached before the magnetic impurities are quenched 
 by the superexchange. 
 While the average of the superexchange interaction,  $<J_{RKKY}> = 0$ 
 vanishes \cite{degennes}, its typical value is of the  
 same order as in a clean sample,
$\sqrt{<J^2_{RKKY}>}= n_M \nu J^2 \cos (2 k_F R)$ \cite{zyuzin} 
 and it has a wide,  
 namely lognormal distribution \cite{lerner}.  
 
 In low dimensions $d \le 2$,  electrons are localised even for 
 weak disorder $g>1$. 
 Recently, it has been argued that in this regime the 
 Kondo renormalisation is stopped by the local level spacing 
 $\Delta_c = 1/\nu_d \xi^d$, where $\nu_d$ is the density of states
 and $\xi$ is the localisation length \cite{meraikh}. 
 Accordingly, it is expected that the distribution 
 of the Kondo temperature converges  to  a finite 
  width and scales with $\Delta_c$, when  size $L$ and 
 the phase coherence length $L_{\varphi}$ exceeds the
 localisation length $\xi$. 
 In general, for finite $g$, corrections to the distribution of the 
 local density of states, and the occurence of anomalously localised
 states will result in further modifications from  the Kondo distribution 
 as  obtained from random matrix theory above.

\section{Conclusions} 

The distribution of the Kondo temperature  depends 
 in a phase coherent metal particle   
 on the time reversal symmetry.  
 The distribution is strongly asymmetric and  scales with 
 the mean level spacing $\Delta$. 
 Its width decrases with increasing 
  concentration of magnetic impurities.

\section*{Acknowledgments} 
The author gratefully acknowledges 
 usefull discussions with  Mikhail Raikh,  Ganpathy Murthy,
 Claudio Chamon,   Peter Fulde, and Eduardo Mucciolo.
 He  acknowledges the  
 hospitality  of the MPI for Physics of Complex Systems,  
 and of the deparment of condensed matter theory of Boston University.
 This research was
supported by  German Research Council (DFG) under  SFB 508, A9.
 and by  EU TMR-network, 
 Grant  HPRN-CT2000-0144.

\end{document}